\documentclass[prl,superscriptaddress,groupaddress,letterpaper,twocolumn,pacs]{revtex4-1}

\usepackage{natbib}
\usepackage{amsthm}         
\usepackage{amsmath} 	   
\usepackage{amssymb}       
\usepackage{amsfonts}
\usepackage{graphicx} 	    
\usepackage{verbatim}
\usepackage{epsfig,bbm}
\usepackage{color}

\begin{document}

\title{Classically stable non-singular cosmological bounces}

\author{Anna Ijjas}
\email{aijjas@princeton.edu}
\affiliation{Princeton Center for Theoretical Science, Princeton University, Princeton, NJ 08544, USA}
\author{Paul J. Steinhardt}
\affiliation{Princeton Center for Theoretical Science, Princeton University, Princeton, NJ 08544, USA}
\affiliation{Department of Physics, Princeton University, Princeton, NJ 08544, USA}

\date{\today}

\begin{abstract}
One of the fundamental questions of theoretical cosmology is whether the universe can undergo a non-singular bounce, {\it i.e.}, smoothly transit from a period of contraction to a period of expansion through violation of the null energy condition (NEC) at energies well below the Planck scale and at finite values of the scale factor such that the entire evolution remains classical. A common claim has been that a non-singular bounce either leads to ghost or gradient instabilities or a cosmological singularity. In this letter, we examine cubic Galileon theories and present a procedure for explicitly constructing examples of a non-singular cosmological bounce without encountering any pathologies and maintaining a sub-luminal sound speed for co-moving curvature modes throughout the NEC violating phase.  We also discuss the relation between our procedure and earlier work.
\end{abstract}

\maketitle

{\it Introduction.} Recently, there has been an increased interest in the question whether it is  possible for the universe to bounce classically, {\it i.e.}, to smoothly transit from a contracting to an expanding phase at a finite value of the scale factor  and at an energy density well below the Planck scale and, hence, far away from a cosmological singularity.  Showing that such classical bounces are possible is a critical step for developing theories of the origin and evolution of the universe that avoid a big bang and its attendant singularity problem, or the invocation of large quantum gravity effects. 

Assuming Einstein gravity and a spatially flat Friedmann-Robertson-Walker (FRW) universe with metric $ds^2= -dt^2+a^2(t)dx^idx_i$ (where $a(t)$ is the scale factor), the challenge of obtaining a non-singular bounce can be best understood by following the evolution of the Hubble parameter $H$, defined as $\dot{a}/a$ (where dot denotes differentiation with respect to time $t$): 
During a period of ordinary (not de Sitter or anti-de Sitter) contraction, $a$ is shrinking while $H$ is becoming more negative and the total energy density $\propto H^2$ is growing. During an ordinary expanding period, on the other hand, $a$ is growing while $H$ is becoming less positive and the total energy density $\propto H^2$ is shrinking. These two cosmological phases can only be connected {\it classically} if, towards the end of the ordinary contracting phase, $H$ reverses its evolution and starts becoming less negative at a finite value of $a$,  well before $H^2$ gets close to Planckian energies. During this `bounce stage,' the increasing value of $H$ eventually hits zero  and continues to grow until it reaches a large positive value (well below the Planck scale but above the nucleosynthesis scale), at which point the bounce stage ends and $H$ begins to decrease.
In a flat FRW universe, a growing Hubble parameter ($\dot{H}>0$), as occurs during the bounce stage, corresponds  to violating the null energy condition (NEC). 

To achieve NEC violation, various forms of stress-energy have been considered \cite{Rubakov:2014jja}.  A common example is a scalar field $\phi$ with higher-order kinetic terms, which can generally lead to pathologies, such as ghost and gradient instabilities. 
To avoid ghost instabilities during NEC violation, it has been a common practice to add higher-order kinetic terms to the action. For example, a simple quartic term $\sim (\partial \phi)^4$ which is characteristic of so-called $P(X)$-theories (where $X=(\partial\phi)^2$) is sufficient to eliminate ghost instabilities; however, NEC violating $P(X)$ theories still suffer from gradient instability \cite{ArkaniHamed:2003uy}.  To alleviate both instabilities,  Galileon-like terms $\sim \Box\phi(\partial \phi)^2$ have been added \cite{Nicolis:2008in}. 

Linear perturbation theory suggests that, for some constructions, Galileon theories can avoid both ghost and gradient instabilities during a period of NEC violation. It has been unclear until now, though, if it is possible for there to be a sufficient period of NEC violation to enable classically stable non-singular Galileon bounces.
In fact, the recent arguments suggest that either the speed of sound of co-moving curvature modes becomes imaginary for some wavelengths during the NEC violating phase \cite{Cai:2012va,Battarra:2014tga} or the evolution must reach a singularity \cite{Libanov:2016kfc}.
 
\begin{figure}[!t]
\includegraphics[width=8cm]{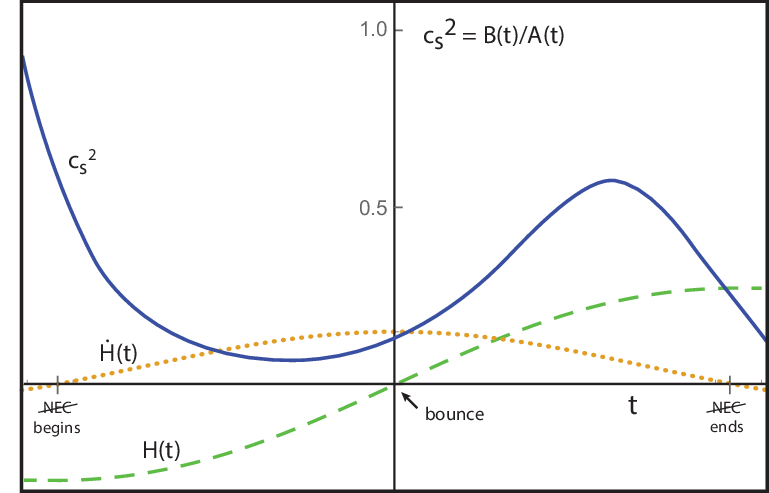}
\caption{A plot of the sound speed $c_S^2$ (solid blue curve) for co-moving curvature perturbations as a function of time $t$. The time coordinate is given in Planck units and the value of $c_s^2$ is given in units where the speed of light is unity.  Superimposed for illustration purposes are the shapes of the background solutions for $H(t)$ (dashed green curve; also shown in Figs.~\ref{fig2} and~\ref{fig3}) and $\dot{H}(t)$ (dotted red curve).   More specifically, the results correspond to  $H(t)=H_0t\, e^{-\alpha(t-t^{\ast})^2}$  and $\gamma(t)= \gamma_0 e^{3 \Theta t } + H(t)$ with the parameter values $H_0=3\times10^{-5}, t^{\ast}=0.5, \alpha=9\times10^{-5}, \gamma_0=-0.0044, \Theta=0.0046$. Notably, throughout, the sound speed is real ($A(t),\, B(t)>0$) and sub-luminal, with $0<c_S^2<1$.
The characteristic energy scale $\sim H^2$ is well below the Planck scale, and the NEC violating phase lasts $\sim 150$ Planck times; it starts when $\dot{H}$ becomes positive at $t_{\rm beg}\simeq-74 M_{\rm Pl}^{-1}$ and ends when $\dot{H}$ becomes negative at $t_{\rm end}\simeq75 M_{\rm Pl}^{-1}$; the bounce ($H(t)=0$) occurs at $t=0$. Note that the bounce stage occurs well within the classical regime.}
\label{fig1}
\end{figure}

\begin{figure}[!tb]
\includegraphics[height=5.5cm]{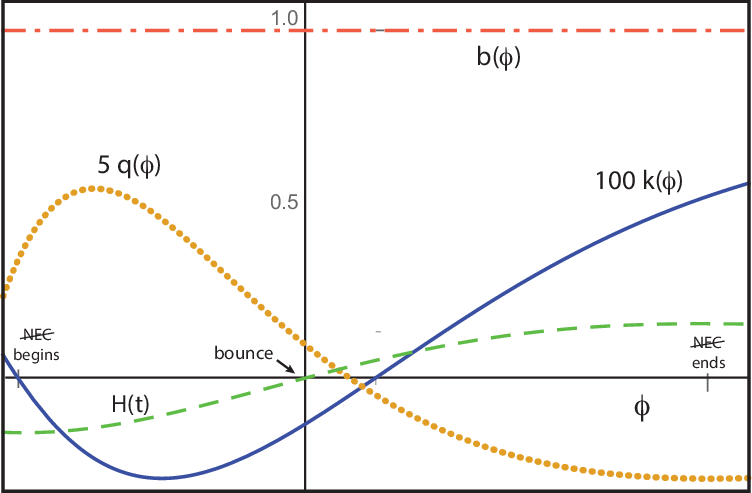}
\caption{A plot of the dimensionless couplings $k$ (solid blue curve), $q$ (dotted orange curve), and  $b$ (dot-dashed red line) as a function of $\phi$ for the example given in Fig.~\ref{fig1}. The $x$ axis has Planck units and the $y$ axis has dimensionless units.  As indicated by the labeling, we have rescaled the quadratic and quartic couplings for the purpose of illustration.}
\label{fig2}
\end{figure}

In this letter, we demonstrate that a classically stable non-singular bounce is possible.  As an example, we study generalized cubic Galileon theories and introduce an `inverse method'  for explicitly constructing examples of classically stable non-singular bounces.  
After briefly reviewing the background evolution during the bounce stage, we derive the second-order Galileon action in co-moving gauge and formulate the linear stability criteria for gauge-invariant curvature perturbations. We then introduce our inverse method and describe how to produce examples that have no ghost or gradient instability and maintain a sub-luminal sound speed throughout the NEC violating phase. Those more interested in the existence of stable bouncing solutions than in the method for obtaining them may wish to jump ahead to Figs.~\ref{fig1} and~\ref{fig2} to see an explicit example, Finally, we discuss the relation between our results and earlier work.

{\it Background evolution during the bounce stage.} We assume that the bounce stage is driven by a single scalar field $\phi$ that is described by the generalized cubic Galileon action $S = \int d^4x \sqrt{-g}\, {\cal L}$ with the defining Lagrangian density
\begin{eqnarray}
\label{FullLagrangian}
{\cal L} &= &
\frac{1}{2} M_{\rm Pl}^2R -\frac{1}{2}k(\phi)(\partial\phi)^2 + \frac{1}{4}M_{\rm Pl}^{-4}q(\phi)(\partial\phi)^4 +
\nonumber\\
&+&    \frac{1}{2}M_{\rm Pl}^{-3}b(\phi)(\partial\phi)^2\Box\phi - V(\phi)
\,.
\end{eqnarray} 
Here, $M_{\rm Pl}$ is the reduced Planck mass; $R$ is the Ricci scalar; $g$ is the metric determinant; $k(\phi)$ is the dimensionless quadratic coupling and $q(\phi)$ is the dimensionless quartic coupling;  $b(\phi)$ is the dimensionless coupling of the scalar field $\phi$ to the cubic Galileon term; and $V(\phi)$ is the scalar potential. Other energy components, such as radiation, matter, dark energy, or other scalars that drive different stages of cosmic evolution are  subdominant and, hence, negligible during the bounce stage  (though they play an important role after the bounce stage, as we will describe in the Discussion section). 

Varying the action with respect to the metric $g_{\mu\nu}$, we find the corresponding Friedmann equations for a spatially flat geometry, 
\begin{eqnarray}
\label{EF1-FRW}
3 H^2  &=& \rho=\frac{1}{2}k\dot{\phi}^2 +  \frac{1}{4}\left(3 q -2b'\right)\dot{\phi}^4 +  3Hb\dot{\phi}^3 +V
\,,\quad \\
\label{EF2-FRW}
-2\dot{H} &=& \rho+p =k\dot{\phi}^2  +  \left(q-b'\right) \dot{\phi}^4 + 3Hb \dot{\phi}^3 -     b\ddot{\phi}\dot{\phi}^2   
,\quad
\end{eqnarray}
where prime denotes differentiation with respect to the scalar field $\phi$. Throughout, we work in reduced Planck units ($M_{\rm Pl}\equiv 1$). The first Friedmann equation describes the different contributions to the total energy density  $\rho$ while the second Friedmann equation describes the sum of energy density and pressure $p$ of the scalar field $\phi$. The ratio $-\dot{H}/H^2$ is equal to the equation-of-state parameter $\epsilon\equiv (3/2)(\rho+p)/\rho$.

Variation of the action with respect to $\phi$ yields the FRW scalar field equation
\begin{eqnarray}
\label{EF3-FRW}
&&\left( k+ \left(3 q -2b'\right) \dot{\phi}^2 +   6 Hb\dot{\phi} + \frac{3}{2}b^2\dot{\phi}^4 \right)\ddot{\phi} =
\nonumber\\
&&= - \frac{1}{2}k'\dot{\phi}^2 -  \frac{ 1}{4}\left(3q' -2b''\right)\dot{\phi}^4
-  \frac{3}{4}b  \left( q\dot{\phi}^4 + 4V \right)\dot{\phi}^2
\nonumber \\
&&
-  \left(k + q \dot{\phi}^2 +  \frac{3}{2}b^2\dot{\phi}^4 
\right) 3 H\dot{\phi}  
 -V'
. 
\end{eqnarray}
The universe enters the bounce stage when the Galileon field's kinetic energy becomes the dominant energy component and the sum of pressure and energy density turns negative (NEC violation). Since $\rho\geq 0$ throughout, the bounce stage is characterized by negative pressure $-p>\rho$ and a super-stiff equation of state $\epsilon <0$ both commonly associated with potential ghost or gradient instabilities. To understand why NEC violating theories are potentially unstable  and under which conditions instabilities can be avoided,  we next derive the stability criteria for Galileon theories.

{\it Stability criteria from linear perturbation theory.}
On a homogeneous FRW background with $\rho\geq 0$, any leading-order instability comes from the kinetic or gradient terms of the linear theory. Hence, to properly identify the stability behavior, we will study first-order perturbations around the smooth background given by Eqs.~(\ref{EF1-FRW}-\ref{EF3-FRW}). 

To perform the stability analysis, it proves useful to employ the ADM formalism and decompose the metric as
\begin{equation}
ds^2 = -\left(N^2 - N^i N_i \right)dt^2 + 2N_i  dx^idt + g_{ij}dx^i dx^j,
\end{equation}
where $N$ is the lapse, $N_i$ is the shift, and $g_{ij}$ is the spatial metric. For the homogeneous FRW background,   $\bar{N}=1$, $\bar{N}_i=0$, and $\bar{g}_{ij}=a^2(t)\delta_{ij}$ (bar denotes background quantities).
We introduce linear perturbations to $N$, and $N_i$ using the standard parametrization
\begin{equation}
\delta N = N-\bar{N} =N_1,\quad
\delta N_i = N_i-\bar{N}_i = \partial_i \chi
. 
\end{equation}

For scalar perturbations of the spatial metric, we have the freedom to choose a particular slicing of space-time, {\it i.e.}, fix the gauge. 
We choose to work in co-moving gauge, where space-time is sliced such that spatial inhomogeneities are all promoted to the metric and the scalar field does not carry any perturbations,
\begin{equation}
\delta\phi = 0,\quad g_{ij}=a^2(t)\left(1+2\zeta(t, {\bf x})\right)\delta_{ij}\,.
\end{equation}
This gauge has the advantage that the co-moving curvature perturbation $\zeta$ is gauge invariant and, hence, ensures that our conclusions about stability do not entail gauge artifacts. In addition, the lack of scalar-field perturbations  significantly simplifies the calculation in the presence of higher-order kinetic terms.

In co-moving gauge, the second-order action takes the form
\begin{equation}
\label{comoving-expanded}
S^{(2)}_{\zeta} = \frac{1}{2}\int d^4x \,a^3 {\cal L}^{(2)}_{\zeta}
\end{equation}
with the Lagrangian density
\begin{eqnarray}
\label{comoving-expanded-lagrangian}
{\cal L}^{(2)}_{\zeta} &=  &  N_1^2 \left( - 6H^2+ k\dot{\phi}^2 + (3q-2b') \dot{\phi}^4 +12Hb\dot{\phi}^3 \right) 
\\
 &+& 4N_1 \left( 3\left(H-\frac{1}{2}b\dot{\phi}^3\right)\dot{\zeta} 
 - \frac{\Delta\zeta}{a^2} 
 \right )  - 6\dot{\zeta}^2 + 2\left(\frac{\nabla \zeta}{a}\right)^2 
\nonumber\\
 &+& 4\frac{\Delta\chi}{a^2}\left( \dot{\zeta} - \left(H- \frac{1}{2}b\dot{\phi}^3\right) N_1\right) 
 - 6\left(6H-b\dot{\phi}^3\right)\zeta\dot{\zeta}
 \nonumber\\
 &-& 9 \left(3H^2 - \frac{1}{2}k\dot{\phi}^2 - \frac{3q- 2b'}{12}\dot{\phi}^4 -Hb\dot{\phi}^3 + V\right)\zeta^2\,,
\nonumber
\end{eqnarray}
where we introduced the spatial gradient $\nabla = \partial_i$ and $\Delta=\nabla^2=\partial_i\partial^i$.
Using the background equations~\eqref{EF1-FRW}~and~\eqref{EF2-FRW} and integrating by parts, the last two terms cancel. Varying the action with respect to the shift, we find the equation of motion for the lapse,
\begin{eqnarray}
\label{shift}
 \left(H - \frac{1}{2}b\dot{\phi}^3\right)N_1 &=& \dot{\zeta}.
\end{eqnarray}
Note that $H - (1/2)b\dot{\phi}^3=0$ renders $\dot{\zeta}=0$. In particular, no singular behavior appears at this point. For our stability analysis, we will consider solutions in which $H - (1/2)b\dot{\phi}^3\neq0$ throughout, and, as we comment below, stability forbids 
 $H - (1/2)b\dot{\phi}^3$ passing continuously through zero.

For $H - (1/2)b\dot{\phi}^3\neq 0$, variation of the second-order action with respect to the lapse $N_1$ and substituting the expression for the shift $\Delta\chi/a^2$ together with the expression for $N_1$ from Eq.~\eqref{shift} back into the original action in Eq.~\eqref{comoving-expanded} and doing a series of integrations by parts yields the second-order action for co-moving curvature perturbations $\zeta$,
\begin{equation}\label{final-com-act}
S_{\zeta}^{(2)} = \int d^4 x\, a^3(t) \left( A(t)\dot{\zeta}^2 - \frac{B(t)}{a^2(t)}(\nabla\zeta)^2 
\right),
\end{equation}
with the dimensionless coefficients
\begin{eqnarray}
\label{no-ghost-eq}
A(t) &=&  \frac{ k\dot{\phi}^2 +  (3 q-2b') \dot{\phi}^4
+   6 Hb\dot{\phi}^3 + \frac{3}{2}b^2\dot{\phi}^6
   }
{2\left(H - \frac{1}{2}b \dot{\phi}^3 \right)^2}
\, ,\\
\label{no-gradinst-eq}
B(t) &=& \frac{k\dot{\phi}^2 + q \dot{\phi}^4  + 2 b\ddot{\phi} \dot{\phi}^2 + 4 H b\dot{\phi}^3 - \frac{1}{2}  b^2\dot{\phi}^6
}{2 \left(H - \frac{1}{2} b \dot{\phi}^3 \right)^2}\,
.\;
\end{eqnarray}
\\
The conditions for {\it stable} NEC-violation correspond to
 positivity of  $A(t)$  ({\it no-ghost condition}), and
 positivity of  $B(t)$  ({\it no gradient instability}). 
Obviously, in the absence of ghost, the square of the sound speed $c_S^2 = B(t)/A(t) > 0$ if $B(t)>0$. To achieve sub-luminal evolution for co-moving curvature modes, we have to additionally demand that $c_S^2 <1$.

{\it `Inverse method.' }
Using the background equations~(\ref{EF1-FRW}-\ref{EF2-FRW}), it is straightforward to check  that Eq.~\eqref{no-gradinst-eq} can be recast to the simple form 
\begin{equation}
\label{gammaDiffEq}
\frac{{\rm d}}{{\rm d}t}\gamma(t)^{-1} + H(t)\gamma(t)^{-1} = B(t)+1\,,
\end{equation}
where $\gamma(t)= H(t) - (1/2) b(\phi) \dot{\phi}^3(t)$ and $\gamma$ is defined to carry the same dimension as $H$  ($[\gamma]=[H]=M_{\rm Pl}$). 
Eq.~\eqref{gammaDiffEq} is a linear first-order differential equation for $\gamma^{-1}(t)$ with the unique solution 
\begin{equation}
\label{gammaSol}
\gamma(t)= \frac{N(t)}{\gamma_0^{-1} +  \int^t_{t_0} \left(B(t)+1\right)N(t) \,dt} ,\quad \gamma(t_0)=\gamma_0,
\end{equation}
where the auxiliary function $N(t)$ is defined as
\begin{equation}
\label{DefN}
N(t) = \exp\left( \int_{t_0}^t H(t) dt\right)=\frac{a(t)}{a(t_0)}.
\end{equation}
In addition, we can re-express $A(t)$ in Eq.~\eqref{no-ghost-eq} as a function of $H$ and $\gamma$, 
\begin{equation}
\label{A:H-gamma}
A(t) = 3+\frac{ 
6 H^2  + 2\dot{H} + \dot{\gamma}  + 3H\gamma }{\gamma^2}.
\end{equation}
These relations have many uses.  For example, for any choice of $H(t)$ and $B(t)$ we can immediately determine the corresponding $\gamma(t)$ and $A(t)$; or, equivalently, for any choice of background solutions $H(t)$ and $\phi(t)$, we can immediately determine the corresponding $A(t)$ and $B(t)$ and, hence, infer the stability behavior for co-moving curvature modes.  This feature allows us to rapidly search through forms for $H(t)$ and $\phi(t)$ that describe bounces and identify choices for which both $A, B>0$ and the sound speed for co-moving curvature modes is  sub-luminal, as shown in Figure~\ref{fig1}.

It is then straightforward to identify the corresponding couplings in the cubic Galileon Lagrangian, Eq.~\eqref{FullLagrangian}, 
using the background equations~(\ref{EF1-FRW}-\ref{EF2-FRW}):
\begin{eqnarray}
\label{K-H}
k(t)&=&  
-2\left(3 H^2  +2 \dot{H} +\dot{\gamma} +3H\gamma \right)/\dot{\phi}^2(t)   
,\quad\\
\label{q-H}
q(t)  &=& 
\frac{4}{3} \left( 2\dot{H} + \dot{\gamma} + 9 H\gamma  \right)/\dot{\phi}^4(t)
+\frac{2}{3}b'
 \,.
\end{eqnarray}
As we will discuss in the following section, if $\gamma$ is finite and non-zero throughout, we have the freedom to set the Galileon coupling $b\equiv1$. 
Finally, inverting $\phi(t)=\phi_0+\int^t_{t_0} \sqrt[3]{2\left( H-\gamma\right)}\, dt$ and substituting $t(\phi)$ into Eqs.~(\ref{K-H}-\ref{q-H}), we find the expressions for the couplings as a function of $\phi$. 
The three coupling functions corresponding to our example in Fig.~\ref{fig1} are depicted in Fig.~\ref{fig2}.  
Note that both $k(\phi)$ and $q(\phi)$ have simple forms. These can be well-approximated by simple functions of $\phi$:
 we have checked that one can start with these approximate coupling functions and the initial conditions $\phi_0=\phi(t_0), \dot{\phi}_0=\dot{\phi}(t_0)$ to find the background solutions describing a classically stable bounce stage with sub-luminal sound speed for co-moving curvature modes. In sum, we have demonstrated that it is possible for the universe to have a classically stable, NEC violating bounce stage without a singularity or any bad behavior.

{\it Special case: $\gamma\to-\infty$.}
If there is no gradient instability ($B(t)>0$), the strict positivity of the function $N(t)$ defined in Eq.~\eqref{DefN} implies that the integral in the denominator of the expression for $\gamma(t)$ in Eq.~\eqref{gammaSol} is positive definite and increasing monotonically for all $t>t_0$. Hence, for any $\gamma_0<0$, there will be some $\bar{t}>t_0$ such that the denominator reaches zero and $\gamma \to -\infty$. A Taylor series of the denominator about $\bar{t}$ results in a leading linear contribution: $\left(B(\bar{t})+1\right))N(\bar{t})(t-\bar{t})$.  Hence, we see that $\gamma$ must approach $-\infty$ as $t \rightarrow \bar{t}$ from $t<\bar{t}$ and $+\infty$ as $t \rightarrow \bar{t}$ from $t>\bar{t}$.  In the previous section, we have shown that $\gamma$ can be chosen such that it is negative and finite  throughout the bounce stage, which means that, formally, $\bar{t}$ would be reached after the NEC is restored, by which time other contributions to the stress-energy (such as matter and radiation) that could be ignored during the bounce stage may dominate and the formal calculation of $\bar{t}$ has no actual physical relevance.  

\begin{figure}[!t]
\includegraphics[height=5.5cm]{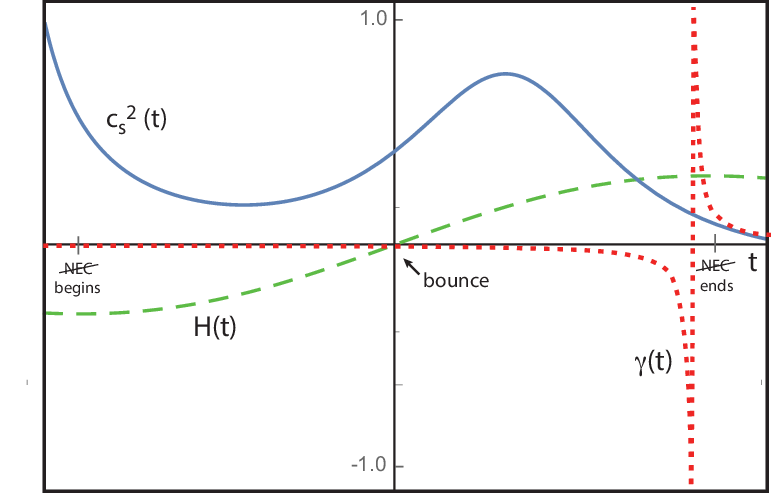}
\caption{A plot of the sound speed $c_S^2$ (solid blue curve) for co-moving curvature modes as a function of time $t$ during the bounce stage  corresponding to the background given by $H(t)=H_0t\, e^{-\alpha(t-t^{\ast})^2}$ (dashed green curve; rescaled to show shape)  for the parameter values $H_0=3\times10^{-5}, t_{\ast}=0.5, \alpha=7\times10^{-5}, \gamma_0=-0.0044$, and $\Theta= 4.6 \times 10^{-6}$, and $\gamma(t)$ as shown by dotted red curve. After the bounce and shortly before the end of NEC violation, $\gamma(t)$ goes from $-\infty$ to $+\infty$ while {\it all} fundamental physical quantities including $H(t)$ and $c_S^2$ remain finite and positive.}
\label{fig3}
\end{figure}

However, in Fig.~\ref{fig3}, we intentionally chose as an academic exercise an example in which $\bar{t}$ occurs during the NEC-violating bounce stage and show that,  even if the derived quantity $\gamma$ diverges during the bounce stage, all fundamental, physical  quantities can remain well-defined and finite throughout the bounce stage.    For this to happen while keeping $H(\bar{t})$, $\dot{\phi}(\bar{t})$ and $\phi(\bar{t})$ finite, it is necessary that the coupling $k(\phi)$, $q(\phi)$ and $b(\phi)$ diverge at $t \rightarrow \bar{t}$. Remarkably, the sound speed $c_S^2$ remains continuous and positive and below 1 despite the diverging coupling.  

It is also possible to construct examples with $\gamma_0>0$ and no gradient instability ($B(t)>0$) throughout the entire bounce stage.  In this case, the denominator in Eq.~\eqref{gammaSol} remains positive definite and monotonically increasing; so the only option is that $\gamma(t)$ monotonically approaches zero from above as $t$ increases, analogous to the examples in Figs.~\ref{fig1} and~\ref{fig2} where $\gamma$ is negative and approaches zero as $t$ decreases.

{\it Discussion.} 
In this Letter, we presented an inverse method that makes it possible to achieve a long-standing goal: to construct field theories with a classically stable, NEC-violating bounce stage that avoids ghost and gradient instabilities, maintains a real, sub-luminal sound speed for co-moving curvature modes, and does not encounter a singularity.   

Our results build on earlier studies on NEC violation and bounces.  In  Refs.~\cite{Deffayet:2010qz,Easson:2011zy}, it was argued that cubic Galileons can produce cosmological bounces and it was shown that there exist stable attractors on FRW cosmological backgrounds, but this work did not analyze the instabilities that can result from curvature fluctuations.   In Ref.~ \cite{Elder:2013gya}, the authors considered Galileon genesis scenarios with NEC violation, and showed that  Galileon scalar field perturbations are stable during the early stages when $H$ is small and gravity is negligible; but this study did not include an exit from the NEC-violating stage when $H$ becomes large and the scalar field can create curvature perturbations with ghost and gradient instabilities.   Indeed, gradient instabilities were encountered during the NEC-violating stage in Refs.~\cite{Cai:2012va,Battarra:2014tga,Qiu:2015nha,Koehn:2015vvy} when the authors attempted to construct non-singular bouncing models, leaving the impression that linear instabilities are unavoidable.  However, this paper definitively shows these instabilities can be safely avoided.

Most recently,  a no-go theorem claiming that singularities are unavoidable in NEC-violating Galileon theories was presented in Ref.~\cite{Libanov:2016kfc} and further generalized in \cite{Kobayashi:2016xpl}. Their argument is equivalent to the statement that $\gamma(t)$ in Eq.~\eqref{gammaSol} must have a zero-point crossing or diverge for some value of $\bar{t}$ for  the Galileon Lagrangian in Eq.~\eqref{FullLagrangian} with $V(\phi)\equiv 0$.  It is important to note, though, that the theorem relies on the assumption that this Lagrangian applies for all times and does not say anything about whether $\bar{t}$ occurs during, before or after the bounce.  We have demonstrated that $\bar{t}$ does not have to occur during or near the NEC-violating bounce stage. As a practical matter, that suffices for constructing non-singular bouncing cosmologies.  Before and after the bounce stage, other forms of energy are non-negligible and even dominant, so the Lagrangian in Eq~\eqref{FullLagrangian} is not to be applicable.   We will present examples in forthcoming work \cite{IjjasSteinhG2}. 

{\it Acknowledgements.} We thank Frans Pretorius and Vasileios Paschalidis  for helpful discussions and comments on the manuscript and we thank Valery Rubakov for stimulating discussions during the PCTS workshop ``Rethinking Cosmology" in May 2016. This research was partially supported by the U.S. Department of Energy under grant number DEFG02-91ER40671.

\bibliographystyle{apsrev}
\bibliography{GalileonBounce}

\end{document}